\pdfoutput=1

\documentclass[journal]{IEEEtran}

\usepackage{amsthm}
\usepackage{amsfonts}
\usepackage{amsmath}
\usepackage{amssymb}
\usepackage{mathrsfs}
\usepackage{graphicx}
\usepackage{subfigure}
\usepackage{stfloats}

\begin{document}
\title{Time-division SQUID multiplexers with reduced sensitivity to external magnetic fields}
\author{G.M.~Stiehl, \thanks{Manuscript received 3 August 2010}%
        H.M.~Cho, \thanks{G.M.~Stiehl, H.M.~Cho, G.C.~Hilton, K.D.~Irwin, J.A.B.~Mates, and C.D.~Reintsema are with the National Institute of Standards and Technology, Boulder, CO 80305 USA (contact G.M.~Stiehl at 303-497-5215 or stiehl@nist.gov)}
        G.C.~Hilton, %
        K.D.~Irwin, %
        J.A.B. Mates, %
        C.D.~Reintsema, \thanks{B.L.~Zink is with the University of Denver, Denver, CO 80210 USA}%
        and~B.L.~Zink \thanks{Contribution of NIST, not subject to copyright}}%

\markboth{Submitted to IEEE transactions on Applied Superconductivity, August 2010, 3EP3J-04}{3EP3J-04}

\maketitle

\begin{abstract}

Time-division SQUID multiplexers are used in many applications that require exquisite control of systematic error. One potential source of systematic error is the pickup of external magnetic fields in the multiplexer. We present measurements of the field sensitivity figure of merit, effective area, for both the first stage and second stage SQUID amplifiers in three NIST SQUID multiplexer designs. These designs include a new variety with improved gradiometry that significantly reduces the effective area of both the first and second stage SQUID amplifiers.

\end{abstract}


\section{Introduction}

\IEEEPARstart{I}{n} recent years, large arrays of superconducting Transition Edge Sensors (TESs) have been implemented in various microcalorimetric and bolometric detector schemes for millimeter and sub-millimeter wavelengths, as well as single-photon detection of optical \cite{Cab00}, x-ray \cite{Ull05} and gamma-ray \cite{Dor07,Zin06} wavelengths. Superconducting QUantum Interference Devices (SQUIDs) are the readout amplifier of choice for TES detectors due to their low noise, low impedance and low power dissipation. Practical readout of TES arrays requires multiplexing at the cold stage to reduce the power load and minimize wiring complexity.  At the National Institute of Standards and Technology (NIST) we have developed various time-domain SQUID multiplexers \cite{Irw04} that have been implemented in bolometric, kilopixel cameras such as the Atacama Cosmology Telescope (ACT) \cite{Fowler10} and the second iteration of the Background Imaging of Cosmic Extragalactic Polarization telescope (BICEP-2) \cite{orlando:77410H}. Experiments such as ACT and BICEP-2 use the SQUID multiplexer designs termed MUX06a and MUX07a. Experiments such as BICEP-2, SPIDER \cite{montroy:62670R} and ACTPol \cite{niemack:77411S} require exquisite control of systematic error in order to resolve the polarization of the Cosmic Microwave Background (CMB). A potential source of systematic error is SQUID pickup from external magnetic fields, which can be synchronous with a telescope's rotation through Earth's magnetic field. We have therefore designed a new SQUID multiplexer, the MUX09a, with the goal of reducing sensitivity to external magnetic fields through improved gradiometry in the SQUID input transformers.

An important figure of merit for SQUID sensitivity to extraneous magnetic fields is the effective area, $A_{eff}=\Phi/B$, where $\Phi$ is the flux coupled to the SQUID by uniform DC magnetic field $B$. The effective area of the first stage SQUID amplifier in the NIST multiplexer is of particular interest, as scan synchronous magnetic pickup from the second stage amplifier can be servoed out \cite{Niemack08}. However, the combined $V$-$\Phi$ response of the first and second stage SQUIDs in the multiplexer designs makes effective area measurements of the first stage SQUID difficult. In this paper, we describe a measurement technique for separating the first and second stage effective areas for SQUID multiplexers. We report effective areas of first and second stage SQUIDs for NIST SQUID multiplexer designs MUX06a, MUX07a and MUX09a. These values show that the improved gradiometry in the MUX09a input transformers has significantly reduced the sensitivity to external magnetic fields.

\section{NIST Time-Division SQUID Multiplexer Designs}

\begin{figure}
\includegraphics[width=3.5in]{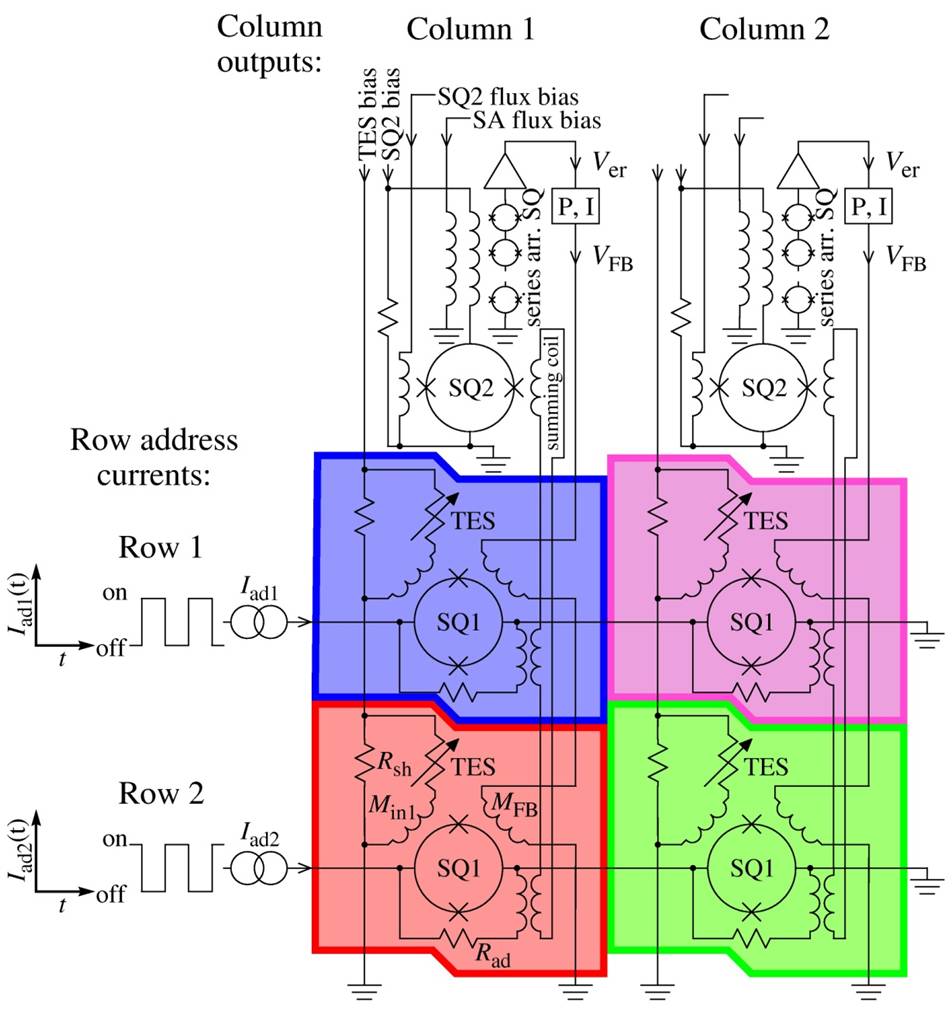}
  \caption{A schematic of a 2x2 subset of the NIST SQUID multiplexer circuit. Each column has two first stage SQUIDs (SQ1) with inputs inductively coupled to a TES. The output of the SQ1s are transformer-coupled to a second stage SQUID (SQ2), which is then read out by a SQUID series array and room-temperature electronics. The four shaded areas represent distinct TES pixels.}
  \label{fig_sim}
\end{figure}

In this section, we briefly describe the NIST time-division SQUID multiplexer circuit and discuss the features that determine effective area contributions in the MUX06a, MUX07a and MUX09a designs.

\subsection{General NIST Time-Division Multiplexer Design}
The NIST time-division SQUID multiplexer circuit consists of 33 first stage SQUIDs (SQ1s), each inductively coupled to a unique TES detector. These SQUIDs are biased sequentially so that only one SQ1 is on at any time. The response from SQ1 couples to the second stage SQUID (SQ2) through a superconducting inductive summing coil that runs the length of the multiplexer chip. This combines the $V$-$\Phi$ response curve of SQ1 and SQ2. The combined SQUID response is further amplified by a SQUID Series Array (SSA) and room-temperature amplifier electronics for readout. A schematic of the multiplexer is shown in Figure \ref{fig_sim}. The NIST time-division SQUID multiplexer circuit is described in greater detail elsewhere \cite{Dor04}.

\subsection{Design Differences: Gradiometry and Input Transformers}

\begin{figure}
\begin{center}
 \includegraphics[width=3.25in]{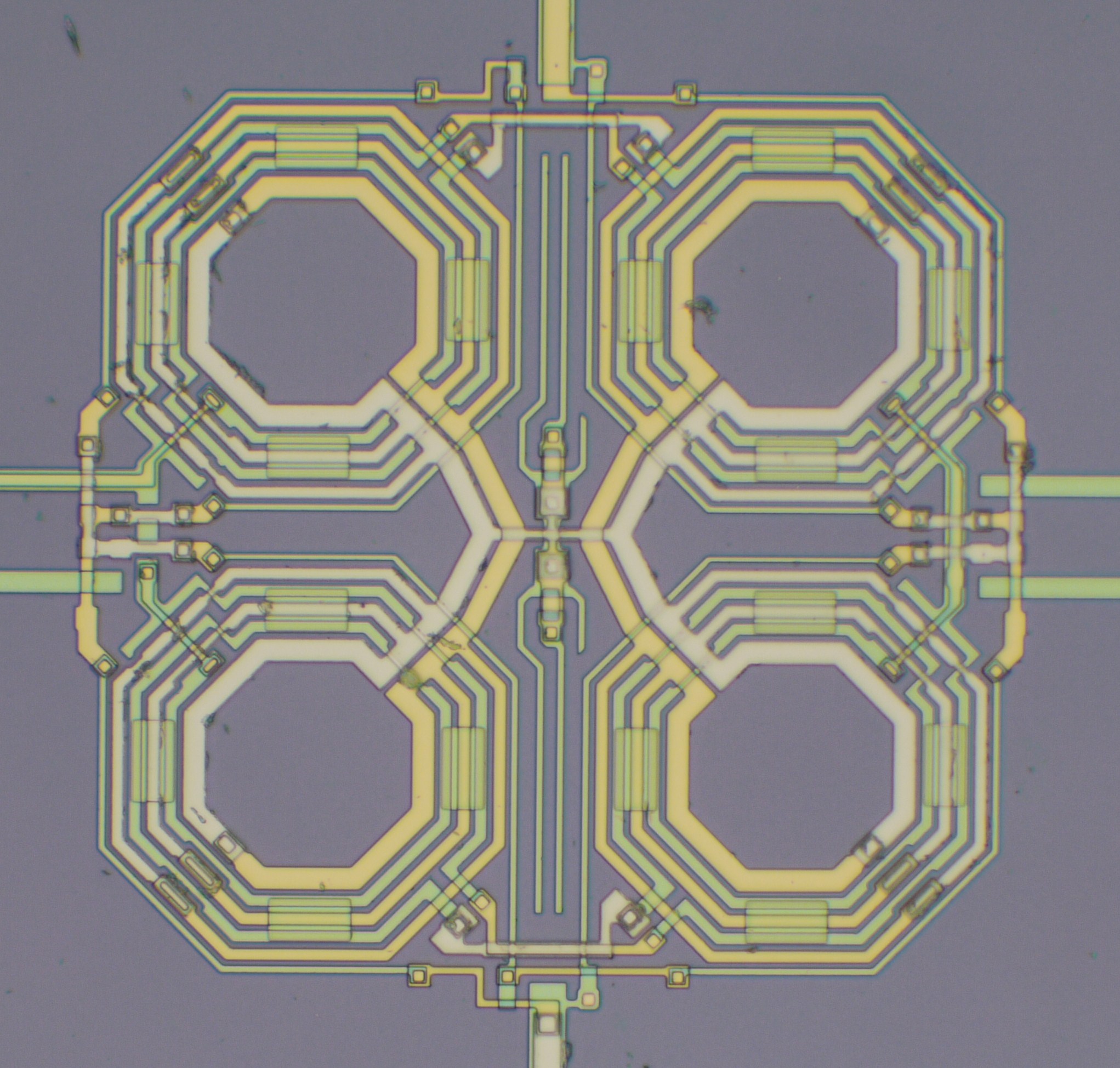}\\
  \caption{A picture of the MUX09a SQUID Gradiometer. For size reference, the inner coil of one of the lobes is 40 $\mu m$ across.}\label{MUX09a_SQ1_top}
  \end{center}
\end{figure}

Since the development of gradiometric SQUIDs at NIST in 1971 \cite{Zimm71}, it has become common practice to use gradiometry in order to reduce a SQUID amplifier's coupling to the external environment. A SQUID gradiometer consists of pickup coils wound in such a way that external magnetic fields do not couple flux into the SQUID. Clover-leaf SQUID gradiometers similar to those developed at PTB \cite{Drung07} are used throughout the NIST multiplexer designs. Figure \ref{MUX09a_SQ1_top} shows the clover-leaf gradiometer used in the MUX09a SQ1.

\begin{figure}
  \includegraphics[width=3.45in]{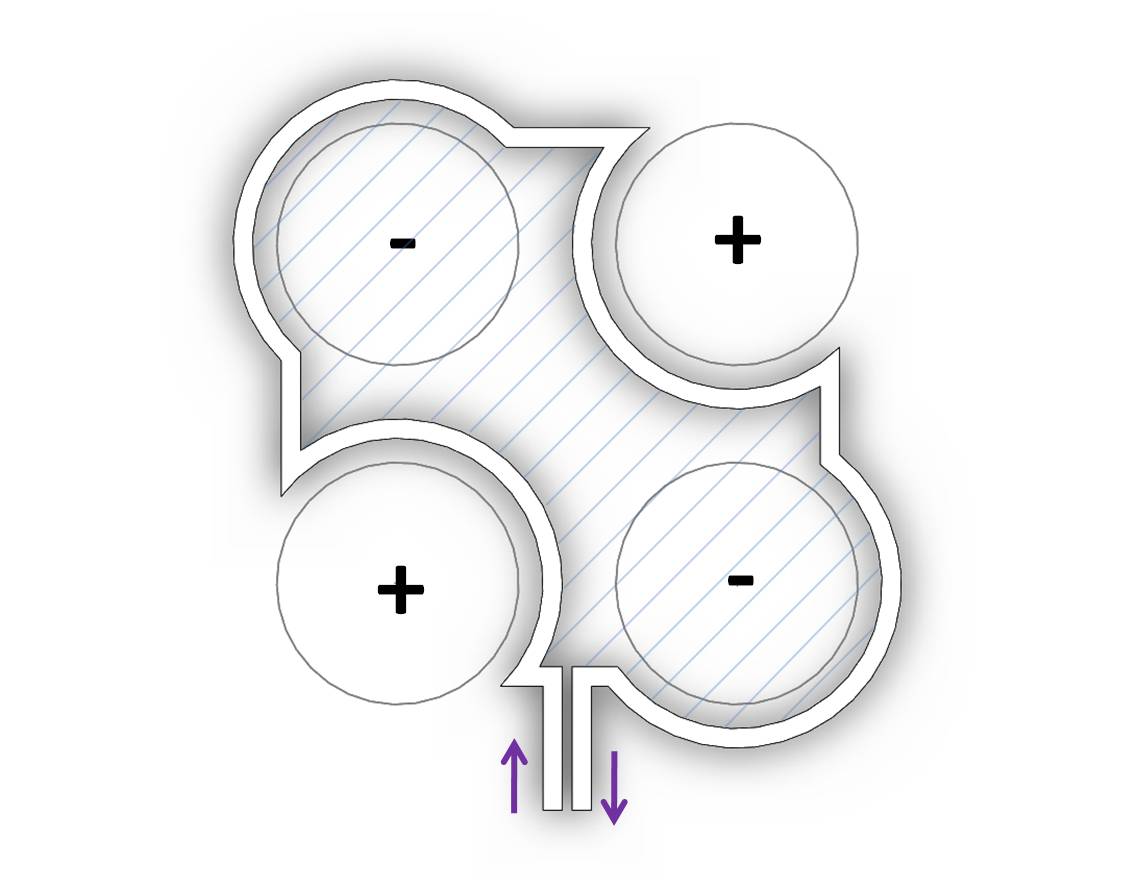}\\
  \caption{A diagram of the half-loop input coil path around the SQUID gradiometer. The arrows show the direction of current in the input coil. The inner circles represent the SQUID gradiometer lobes, and the plus and minus signs give the magnetic coupling polarity. The cross-hatching shows the area susceptible to coupling from external magnetic fields.}\label{HalfLoop}
\end{figure}
\begin{figure}
  \includegraphics[width=3.45in]{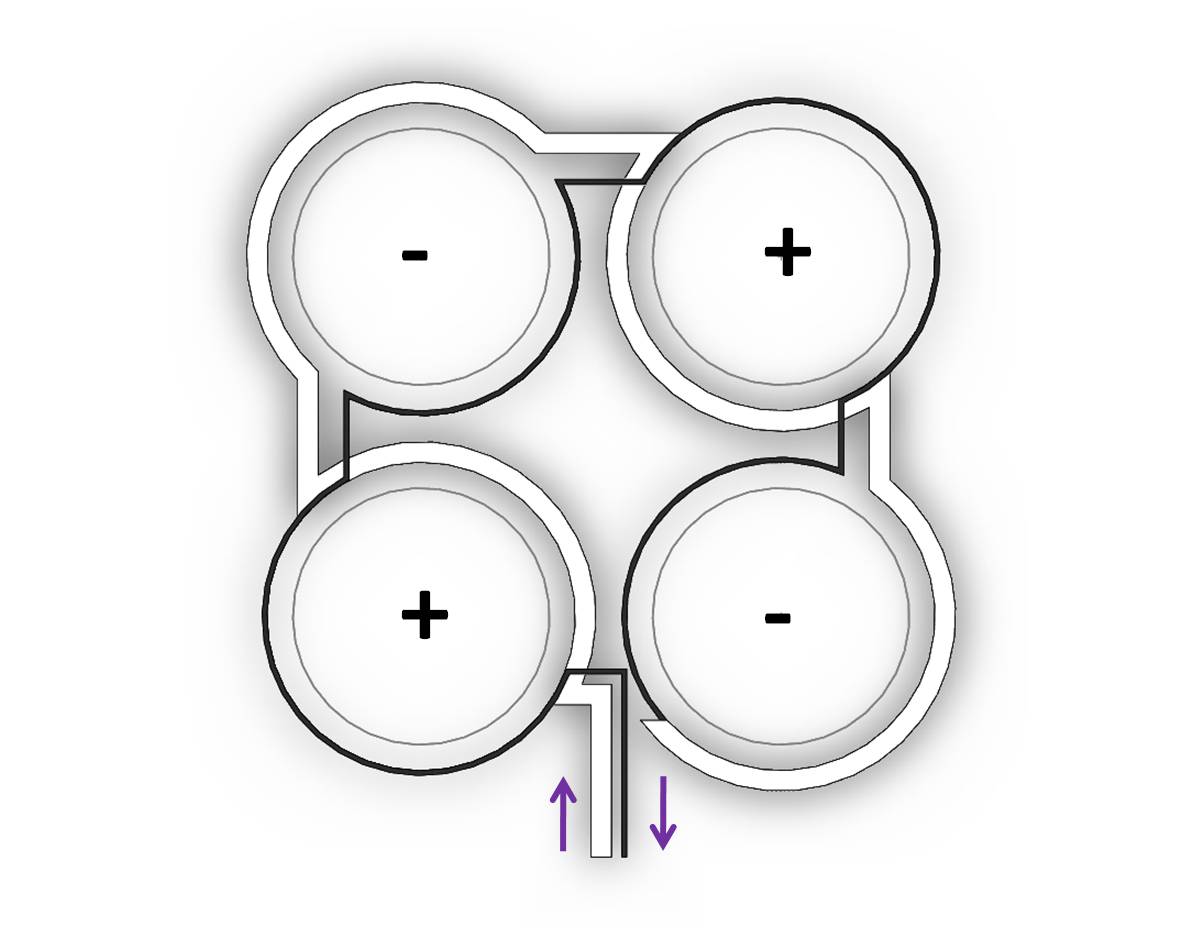}\\
  \caption{A diagram of the whole-loop input coil path around the SQUID gradiometer. The arrows show the direction of current in the input coil. The wider trace is the incoming current path and the thin trace is the return line. The difference in trace size is only for ease of viewing. The inner circles represent the SQUID gradiometer lobes, and the plus and minus signs give the magnetic coupling polarity. Notice how in this configuration flux coupled through the area outlined by the wider trace is canceled by the flux coupled through the area outlined by the thin trace.}\label{WholeLoop}
\end{figure}

In order to couple a signal to the gradiometric SQUID pickup coils, an input coil is wound such that the polarity of the flux coupled into the SQUID matches the polarity of each gradiometer lobe. One such layout is shown in Figure \ref{HalfLoop}. The input coil windings in this figure make half-loop turns of alternating polarity around the SQUID gradiometer lobes in order to couple flux into the SQUID. The MUX06a and MUX07a SQUID multiplexer circuits use half-loop input coil geometries similar to that shown in Figure \ref{HalfLoop}. The MUX06a uses a $4\frac{1}{2}$ turn variation of Figure \ref{HalfLoop} on SQ1, and the MUX07a uses a $1\frac{1}{2}$ turn variation on SQ1. Both designs use a $4\frac{1}{2}$ turn input coil geometry on SQ2. If input coil geometries like that shown in Figure \ref{HalfLoop} are part of a superconducting transformer, screening currents induced in the input coil by an extraneous uniform DC field will couple flux into the SQUID. The new design, the MUX09a, utilizes whole-loop input coil geometries like that shown in Figure \ref{WholeLoop}. The SQ1 has a 2 turn input coil and the SQ2 has a 4 turn input coil. This configuration minimizes coupling to uniform external fields.

The poor gradiometry of the half-loop input coil contributes to the effective area only if the coil is part of a closed superconducting loop, such as a transformer. There is no such input transformer on the first stage SQUID in the MUX06a. Therefore, we do not expect the input coil to couple flux from a uniform DC magnetic field into the SQUID. The MUX06a does however have a superconducting input transformer on the second stage SQUID (the summing coil). Thus, we expect a significant contribution to the effective area of the second stage. The MUX07a design utilizes both the half-loop input coil geometry and input transformers for SQ1 and SQ2. Thus, we expect to see a large effective area for each amplification stage. The MUX09a also has input transformers on SQ1 and SQ2. However, with the whole-loop input coil geometry we do not expect a significant effective area contribution from either amplification stage.

\section{Measurement Scheme}
A liquid helium immersion probe used for SQUID characterization was adapted to make effective area measurements. A solenoid is mounted axially to the cold end of the probe and applies a uniform DC magnetic field perpendicular to the SQUID multiplexer. The solenoid consists of 2100 turns and is 0.2667 meters in length. A high-permeability magnetic shield mounts around the solenoid. The cold end of the probe is submersed in liquid helium.

\begin{figure}
  \includegraphics[width=3.45in]{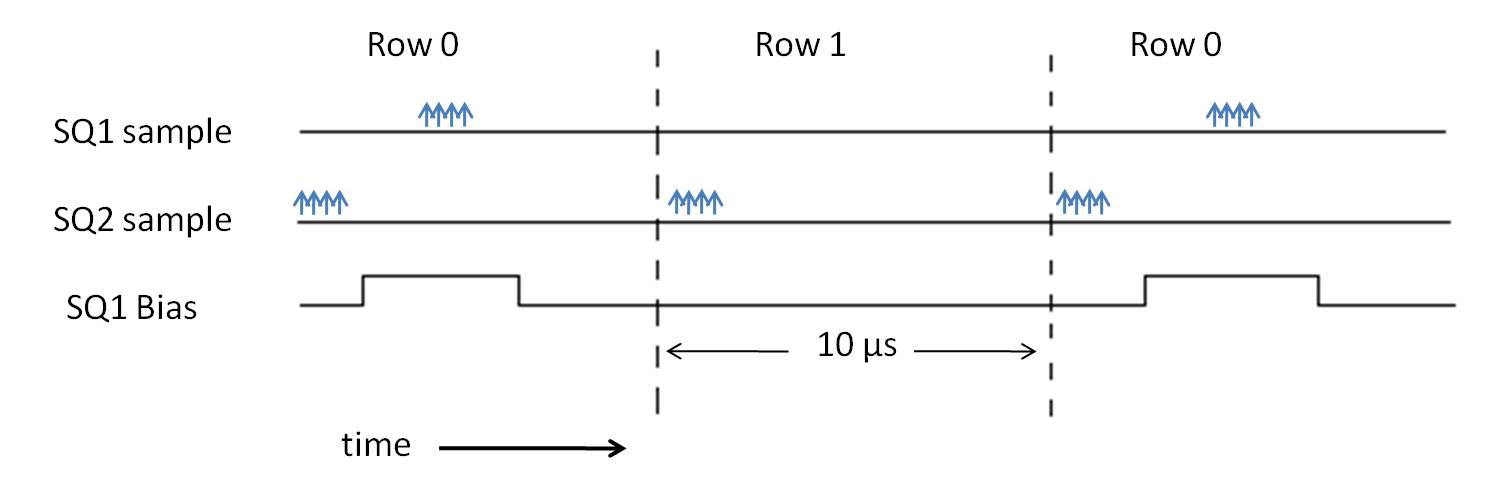}\\
  \caption{A timing diagram of the scheme used in the closed-loop methodology for deconvolving SQ1 and SQ2 effective area contributions. The row period is about 10 microseconds and the arrows depict where the error signal is sampled for the corresponding feedback system.}\label{ClosedLoopScheme}
\end{figure}

As the $V$-$\Phi$ response curves of SQ1 and SQ2 are combined upon readout, measurements of the separate contributions to effective area by the first and second stage SQUID circuits can be cumbersome. We therefore devised a method of separating the effective area contributions through the use of the NIST digital feedback electronics \cite{Rei03} and the multiplexed nested feedback loop scheme shown in Figure \ref{ClosedLoopScheme}.

In this two-row scheme, both SQ1 and SQ2 are closed in flux-locked loops using the same error signal sampled at different times. In the first row of the scheme (row 0), the error signal is sampled and feedback flux is applied to SQ2 prior to SQ1 being turned on. SQ1 is then turned on, the error signal is sampled again and feedback flux is applied to SQ1.

In this configuration, we first measure the flux from the applied external magnetic field coupling into the summing coil and SQ2. The feedback electronics null out the extraneous flux by applying a signal to the SQ2 feedback. The amount of feedback flux required to accomplish this is exactly the amount that is coupled into the SQUID by the magnet. This is the SQ2 contribution to the effective area.

Once the SQ1 is turned on, the flux from the applied field couples into the SQ1. The error signal is sampled again, and feedback is applied to the SQ1 to null the applied flux. This is the SQ1 contribution to the effective area. We thus separate out the SQ1 and SQ2 effective areas.

The solenoid is stepped through 100 different DC bias values. At each solenoid bias many thousands of flux data points are taken and then averaged. The average values for SQ2 feedback and SQ1 feedback are recorded. A linear fit is applied to the feedback flux as a function of applied magnetic field for both  SQ1 and SQ2. The absolute value of the slope from each linear fit determines the effective area.

The magnet current bias values is converted to field values using an equation for the field inside a solenoid:
\begin{equation}
B = \frac{V}{R_{\rm{bias}}}\frac{2100}{0.2667\rm{m}}\mu_{o}
\label{BfieldEQ}
\end{equation}
The magnet calibration was checked with a Gauss meter at higher values of magnetic field. There was no measurable deviation from Equation \ref{BfieldEQ}. The magnet response is thus assumed to be linear at smaller current values.

\section{Results}

\begin{figure}
  \includegraphics[width=3.45in]{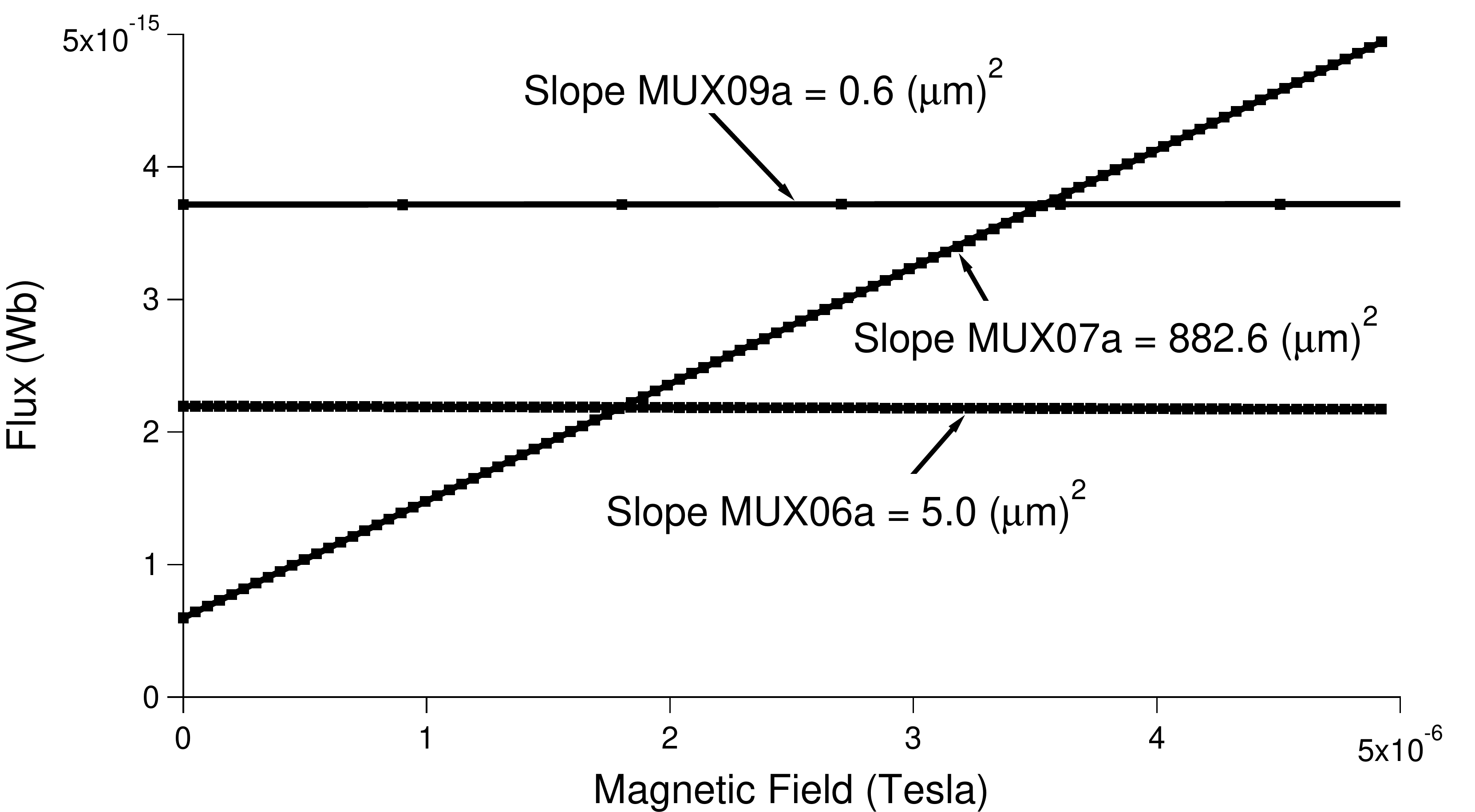}\\
  \caption{Separated effective area contribution of SQ1 for the MUX06a, MUX07a and MUX09a.}\label{SQUID1EffA}
\end{figure}

\begin{figure}
  \includegraphics[width=3.45in]{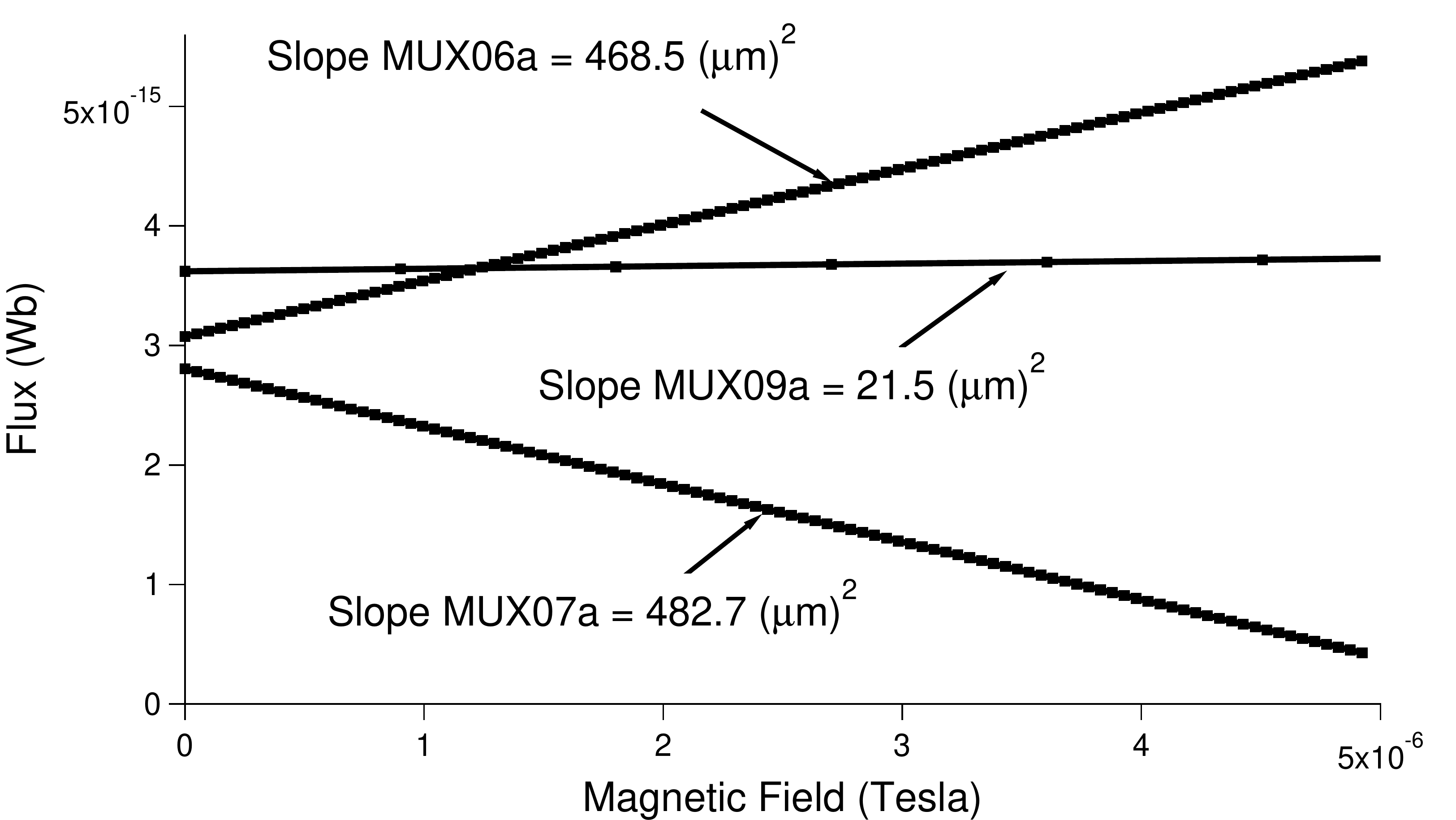}\\
  \caption{Separated effective area contribution of SQ2 for the MUX06a, MUX07a and MUX09a.}\label{SQUID2EffA}
\end{figure}

We performed effective area measurements for SQ1 and SQ2 in the MUX06a, MUX07a and MUX09a. The voltages across the SQ1 feedback and SQ2 feedback lines are recorded and converted into flux as follows:
\begin{equation}
\Phi_{\rm{SQ Fb}} = (V/R_{\rm{FB}})M_{\rm{SQ FB}}
\label{SQ1Feedback}
\end{equation}
where $M_{\rm{SQ Fb}}$ is the feedback mutual inductance and $R_{\rm{FB}}$ is the bias resistor for the SQUID feedback line.

Figures \ref{SQUID1EffA} and \ref{SQUID2EffA} show the amount of flux plotted as a function of applied magnetic field for SQ1 and SQ2 (respectively) for all multiplexer designs. These plots also show a linear fit to the data. The absolute value of the slopes for the linear fits are the effective areas and are tabulated in Table \ref{EffaTable}.

\begin{table}
\caption{Effective Areas for SQUID multiplexers}
\begin{center}
\begin{tabular}{ccccc}
\hline \hline
&  & SQ1 &  & SQ2\\
&  & ($\mu m^2$) &  & ($\mu m^2$)\\ \hline
MUX06a &  & 5.0 &  & 468.5\\
MUX07a &  & 882.6 &  & 482.7\\
MUX09a &  & 0.6 &  & 21.5\\ \hline \hline
\end{tabular}
\label{EffaTable}
\end{center}
\end{table}

A variation in effective area values of a few square micrometers from chip to chip and row to row has been observed. An exhaustive study of this variation and possible row-position dependence is planned but has not yet been conducted.

\section{Conclusion}

We have successfully measured SQ1 and SQ2 effective areas for three NIST SQUID multiplexer designs: MUX06a, MUX07a and MUX09a. From the results, it is clear that the largest contributions to effective area are due to poor gradiometry in the superconducting transformers used in the SQUID multiplexer circuit. Changing the input coil from half-loop to whole-loop gradiometry in the MUX09a multiplexer has made a significant impact by reducing the effective area of SQ2 by factor of 20, and more importantly, by reducing SQ1 by approximately three orders of magnitude when compared to the MUX07a. As a result, the MUX09a is far less sensitive to extraneous magnetic fields than earlier designs. This should significantly reduce the systematic error introduced by pickup from external magnetic fields in instruments that utilize NIST SQUID multiplexers.

\section*{Acknowledgment}
The authors thank D.A. Bennett, W.B. Doriese, R.D. Horansky, G.C. O$'$Neil, M.D. Niemack, and D.R. Schmidt for their insight into SQUID electronics and measurement techniques. We also thank L. Ferreira and T. Sundby for their continued technical support. G.M. Stiehl thanks those involved with the PREP program both at NIST and the University of Denver for their support.

\ifCLASSOPTIONcaptionsoff
  \newpage
\fi

\bibliographystyle{IEEEtran}
\bibliography{Stiehl_IEEE}

\end{document}